\documentclass[aps,prd,nofootinbib,amsmath,amssymb,twocolumn,10pt]{revtex4}
\pdfoutput=1
\usepackage{graphicx}
\usepackage{bm}
\usepackage{dcolumn}
\usepackage{amsmath}
\usepackage{amsmath}
\usepackage{amssymb}

\renewcommand \thesection{\arabic{section}}

\numberwithin{equation}{section}
\usepackage{epsfig}
\begin{document}
\newcommand{\newc}{\newcommand}
\newc{\be}{\begin{equation}}
\newc{\ee}{\end{equation}}
\newc{\ra}{\rightarrow}
\newc{\lra}{\leftrightarrow}
\newc{\lsim}{\buildrel{<}\over{\sim}}
\newc{\gsim}{\buildrel{>}\over{\sim}}
\title{Constraining the Anisotropic Expansion of the Universe}
\author{Rong-Gen Cai$^{1}$}
\email{cairg@itp.ac.cn}
\author{Yin-Zhe Ma$^{2,3}$}
\email{mayinzhe@phas.ubc.ca}
\author{Bo Tang$^{1}$}
\email{tangbo.bit@163.com}
\author{Zhong-Liang Tuo$^{1}$}
\email{tuozhl@itp.ac.cn}
\affiliation{$^{1}$ Key Laboratory of Frontiers in Theoretical Physics, Institute of
Theoretical Physics, Chinese Academy of Sciences, P.O. Box 2735,
Beijing 100190, China.   \\
$^{2}$ Department of Physics and Astronomy, University of British Columbia, Vancouver, V6T 1Z1, BC Canada.   \\
$^{3}$ Canadian Institute for Theoretical Astrophysics, Toronto, M5S 3H8, Ontario, Canada.}
\date{\today}

\begin{abstract}
We study the possibly existing anisotropy in the accelerating
expansion universe with the Union2 Type Ia supernovae data and
Gamma-ray burst data. We construct a direction-dependent dark
energy model and constrain the anisotropy direction and strength
of modulation. We find that the maximum anisotropic deviation
direction is $(l,\,b)=(126^{\circ},\,13^{\circ})$ (or equivalently
$(l,\,b)=(306^{\circ},\,-13^{\circ})$), and the current anisotropy
level is $g_0=0.030_{+0.010}^{-0.030}$ ($1\sigma$ confidence level
with Union2 data). Our results do not show strong evidence for the
anisotropic dark energy model. We also discuss potential methods
that may distinguish the peculiar velocity field from the
anisotropic dark energy model.
\end{abstract}

\maketitle
\section{\normalsize{Introduction}}
\renewcommand{\thesection}{\arabic{section}}

The cosmological principle has played an important role in modern
cosmology~\cite{Weinberg}. It tells us that our universe is
homogeneous and isotropic on large cosmic scale, which is
consistent with currently observational data sets such as the
cosmic microwave background (CMB) radiation data from the
Wilkinson Microwave Anisotropy Probe
(\textit{WMAP})~\cite{WMAP7,Bennett:2010jb,Hinshaw:2012fq} and
\textit{Planck} satellite~\cite{Planck2013-1}.
Up to now, current astronomical observations are still in good
agreement with $\Lambda$CDM model generally~\cite{LCDM}.

Despite the fact that the concordance cosmological model
($\Lambda$CDM model) is confirmed by many observational data, the
model is also challenged by some
observations~\cite{challenge,Mariano:2012ia} (see~\cite{lcdm} and
references therein for more details). Recently, Appleby
and Linder~\cite{Appleby:2012as} found that an anisotropic dark
energy model that preserves isotropic expansion to the level
required by  CMB still needs  to be further considered. Therefore,
it is important and necessary to check the cosmological principle
with current available observational data. As more Type Ia
supernovae (SNIa) data~\cite{Union2, Union2.1} and high-redshift
Gamma-ray burst (GRB) data are
released~\cite{Wei:2010wu,Liang:2008kx}, it becomes possible to
detect the anisotropic direction of cosmic expansion by using the
supernovae data and the GRB data.

Indeed, a lot of effects may cause the cosmic anisotropy.
For example, peculiar velocities may lead observers to find that
the observed acceleration is maximized in one direction but
minimized in the opposite~\cite{peculiar}. A vector field model of
dark energy  may lead to a direction-dependent equation of
state~\cite{vector}.
Many data analyses have already been made to search for the cosmic anisotropy.
Using Union2 of Type Ia supernovae (SNIa) data,
Ref.~\cite{noevidence} derived the angular covariance function of
the standard candle magnitude fluctuations, and the authors  did
not find any angular scales where the covariance function deviates
from $0$ in a statistically significant manner. By using $288$
SNIa~\cite{Kessler:2009ys}, Davis {\it et al.}~\cite{davis}
studied the effects of peculiar velocities, taking into
consideration of our own peculiar motion, supernova's motion and
coherent bulk motion, and found that it can cause a systematic
shift $\Delta w=0.02$ in the equation of state of dark energy if
one neglected coherent velocities. Gupta {\it et
al.}~\cite{previous} introduced a statistics based on the extreme
value theory and applied it to the gold data set of SNIa, and they
showed that the data is consistent with isotropy and Gaussianity.
Cooray {\it et al.}~\cite{Cooray:2008qn} used the SNIa to probe
the spatial inhomogeneity and anisotropy of dark energy, showing
that a shallow, almost all-sky survey can limit the \textit{rms}
dark energy fluctuations at the horizon scale down to a fractional
energy density of $\sim10^{-4}$. On the other hand,  a
``residual'' statistics was constructed in \cite{previous2} to
search for the preferred direction in different slices of past
light-cone, the authors  found that at low redshift ($z<0.5$) an
isotropic model was not consistent with the SNIa data  even at
$2$-$3\sigma$. Campanelli {\it et al.}~\cite{previous1} found that
anisotropy is permitted both in the geometry of the universe and
in the dark energy equation of state, if one worked in the
framework of an anisotropic Bianchi type I cosmological model and
the presence of an anisotropic dark energy equation of state.
Furthermore, Refs.~\cite{schwarz,challenge1,Cai:2011xs} used the
hemisphere comparison method to fit the $\Lambda$CDM model (and
$w$CDM model) to the supernovae data, and detected a preferred
axis at statistically significant level. These results are
consistent with many other observations, such as the CMB
dipole~\cite{cmbdipole}, large scale alignment in the QSO optical
polarization data~\cite{quasar} and large scale velocity
flows~\cite{velocity}. Ref.~\cite{challenge1} obtained the average
direction of the preferred axes as $(l,\,b)=(278^{\circ}\pm
26^{\circ},\,45^{\circ}\pm 27^{\circ})$. Further analysis was made
by~\cite{Kalus:2012zu}, where the authors used different
low-redshift ($z<0.2$) SNIa samples and employed the the Hubble
parameter to quantify the anisotropy level, and the results showed
that all the SNIa samples indicated an anisotropic direction at
95\% confidence level. Finally, we should mention that an
anisotropic universe model or anisotropic dark energy model can
potentially solve the CMB low-quadrupole
problem~\cite{Rodrigues:2007ny,Beltran
Jimenez:2007ai,Campanelli:2006vb,Planck22}.

In this paper we study the plausible anisotropy in the accelerated
expanding Universe with the Union2 data. We construct an
anisotropic dark energy model and aim to detect the maximum
anisotropy direction that deviates from the isotropic dark energy
model described by $\Lambda$CDM model. Furthermore, we consider
the impact of redshift on the direction by using the redshift
tomography method, with the high-redshift Gamma-ray burst data as
a complement data set. Finally, we check two other models as the
description of the isotropic background. One is the $w$CDM model,
the other is a dynamical dark energy model represented by the
Chevallier-Polarski-Linder (CPL) parametrization~\cite{CPL}. We
exam that if our results are dependent on the isotropic
background.

The paper is organized as follows. In the next section we give a
general introduction to the anisotropic dark energy model, which
is based on the isotropic background described by the $\Lambda$CDM
model, and the $\chi^2$ statistics of the model with observational
data. In Sec. 3, we give the numerical results on the maximum
anisotropy directions from the SNIa data and high-redshift GRB
data with different slices of redshift. We also give the results
for the $w$CDM and CPL model. Our conclusions are presented in Sec.4.


\section{\normalsize{Anisotropic Dark Energy Model}}
\renewcommand{\thesection}{\arabic{section}}

If dark energy has anisotropic repulsive force, it will directly
affect the expansion rate of the universe, leading to the
anisotropic luminosity distance. This effect should be observable
by the luminosity of SNIa. In this paper, we use the Union2 data
set~\cite{Union2}, which contains $557$ SNIa data covering the
redshift range $z=[0.015,\,1.4]$. In addition, we
incorporate a GRB data set with $67$ GRB samples up to $z=6.6$
(see Table~\ref{tab:GRB}).

We try to quantify the deviation from the isotropic background as dipole modulation. By using the luminosity distance we define the deviation
from isotropic expansion as
\begin{equation}
\frac{d_{\rm L}(\vec{z})-d_{\rm L}^0(z)}{d_{\rm
L}^0(z)}=g(z)(\hat{z}\cdot \hat{n}). \label{deviation}
\end{equation}
where, the true luminosity distance of the supernova is $d_{\rm
L}(\vec{z})$, and in an isotropic background, the luminosity
distance is $d_{\rm L}^0(z)$. $g(z)(\hat{z}\cdot \hat{n})$ is the
modulation part of the luminosity distance, which makes the real
luminosity function anisotropic. Note that the modulation could be
any power-law form of $(\hat{z}\cdot\hat{n})$, such as
$(\hat{z}\cdot\hat{n})^s$, where $s$ is a constant, but we focus
on the dipole modulation here ($s=1$). In principle, one can use
this model to discuss quadrupole modulation ($s=2$), octupole
modulation ($s=4$) and higher moments ($s>4$). $\hat{z}$ is the
unit direction vector of the supernova, which can be expressed by
using the Galactic coordinate system. $\hat{n}$ is the direction
of dark energy dipole, which is the maximal expanding direction,
\begin{equation}
\hat{n}=(\cos\phi \sin\theta,\sin\phi\sin\theta,\cos\theta )
\end{equation}
where $\theta\in [0, \pi)$ and $\phi\in [0, 2\pi)$. For
the modulation of strength $g(z)$, one can consider the simplest
case, $g(z)=g_0$, which corresponds to the case where the
direction-dependent modulation is constant over all redshifts. Of
course, we can also parameterize $g(z)$ with linear function of
$z$ as
\begin{equation}
g(z)=g_0+g_1z.
\end{equation}
where $g_0$ and $g_1$ are two constants, representing the strength
of modulation and the time evolution of modulation, respectively.
By parameterizing $g(z)$ with linear function, one can detect the
redshift dependence of the anisotropy.
Here $g_0$ represents the redshift independent part of the deviation and $g_1z$  stands for  the redshift linearly dependent
 part of the deviation. Of course, in principle, we can include higher
order expansion terms of the Taylor expansion of the modulation
function, however, including those terms  will introduce more free
parameters and reduce the constraining ability on the parameters.
Therefore we limit to the case with the linear term here.
Note that here the linear function is only a representative of various parametrization forms, and it might be invalid at high redshifts. To overcome
this, one may take other forms of parametrization, for example, a CPL-like  parametrization, $g(z)=g_0+g_1\frac{z}{1+z}$. In this case, the divergence will
not appear when $z \to \infty$. Since we mainly focus on the SNIa data with the highest redshift $z=1.4$, in this
paper we therefore consider two cases with modulation function as
$g(z)=g_0$ and $g(z)=g_0 +g_1z$, respectively. We do not expect other
parametrization forms will change significantly our main conclusions.

In a spatially flat isotropic cosmological background, the luminosity distance can be expressed as
\begin{equation}
d_{\rm L}^0(z)=(1+z)\int_{0}^{z}\frac{H_0}{H({z}')}d{z}'.
\label{distance}
\end{equation}
where $H_{0}=100h~\textrm{km} \, \textrm{s}^{-1} \,
\textrm{Mpc}^{-1}$ is the Hubble constant. Accordingly, the
theoretical distance modulus $\mu_{{\rm th}}$ is defined as \be
\mu_{{\rm th}}(z)=5\log_{10}d_{\rm L}(z)+\mu_{0}~,~~\mu_{0}=42.384-5\log_{10}h~.\nonumber\\
\ee

Since $\Lambda$CDM model is consistent with current astronomical
observations, it is reasonable to take $\Lambda$CDM model as the
isotropic base model, in which the Hubble parameter can be
expressed as
\begin{equation}
H^2(z)=H^2_0[\Omega_{\textrm{m0}}(1+z)^3+(1-\Omega_{\textrm{m0}})].
\end{equation}
where $\Omega_{\textrm{m0}}$ is the current value of the energy
density fraction of matter. While in the case of $w$CDM model, the
equation of state of dark energy is parametrized by a constant
$w$, $w=p/\rho$, therefore we have
\begin{equation}
H^2(z)=H^2_0[\Omega_{\textrm{m0}}(1+z)^3+(1-\Omega_{\textrm{m0}})(1+z)^{3+3w}].
\end{equation}
And if the background is described by the CPL parametrization, the
equation of state of dark energy is $w=w_0+w_1\frac{z}{1+z}$.
Accordingly one obtains
\begin{eqnarray}
H(z)&=& H_0^2[\Omega_{\textrm{m0}}(1+z)^3 +
(1-\Omega_{\textrm{m0}})(1+z)^{3(1+w_0+w_1)} \nonumber \\
&\times& \exp(-3w_1z/(1+z))].
\end{eqnarray}

We employ the Union2 data set and the GRB data to constrain the
anisotropic dark energy model. The directions to the SNIa we use
here are given in Ref.~\cite{noevidence}, and are described in the
equatorial coordinates (right ascension and declination).
The 67 GRB data are shown in Table~\ref{tab:GRB}. These
samples are selected from~\cite{Wei:2010wu,Liang:2008kx} and we
add in the position of each data point \footnote{For more details,
please visit the website
http://www.mpe.mpg.de/~jcg/grbgen.html.}. In order to make
comparisons with other results, we convert these coordinates to
the galactic coordinates $(l,b)$~\cite{convert}.

Let us suppose the experiment error between each measurement is completely independent, so
the covariance matrix can be simplified as the diagonal component, and the \textbf{$\chi^{2}$}
can be written as
\[
\chi^2=\sum_{i=1}^{N}\frac{[\mu_{\rm obs}(z_{i})-\mu_{\rm
th}(\vec{z_{i}})]^{2}}{\sigma^{2}(z_{i})},
\]
where $N$ is the number of data ($N=557$ for SNIa, $N=67$ for GRB,
thus $N=624$ for combined data set). $\mu_{\rm obs}(z_{i})$ is the
measured distance modulus from the data, and $\mu_{\rm th}(z_{i})$
is the direction-dependent theoretical distance modulus.

We can eliminate the nuisance parameter $\mu_{0}$ by expanding \textbf{$\chi^{2}$} with
respect to $\mu_{0}$ \cite{Nesseris:2005ur}:
\begin{equation}
\chi^{2}=A+2B\mu_{0}+C\mu_{0}^{2},\label{eq:expand}\end{equation}
 where \begin{eqnarray*}
A & = & \sum_{i}\frac{[\mu_{{\rm th}}(z_{i};\mu_{0}=0)-\mu_{{\rm obs}}(z_{i})]^{2}}{\sigma^{2}(z_{i})},\\
B & = & \sum_{i}\frac{\mu_{{\rm th}}(z_{i};\mu_{0}=0)-\mu_{{\rm obs}}(z_{i})}{\sigma^{2}(z_{i})},\\
C & = & \sum_{i}\frac{1}{\sigma^{2}(z_{i})}.\end{eqnarray*}
The \textbf{$\chi^{2}$} has a minimum as
\begin{equation}
\tilde{\chi}^{2}=A-B^2/C~,\label{sn}
 \end{equation}
which is
independent of $\mu_{0}$. This technique is equivalent to performing a
uniform marginalization over $\mu_{0}$~\cite{Nesseris:2005ur}. We will adopt
\textbf{$\widetilde{\chi}^{2}$} as the goodness of fitting instead of \textbf{$\chi^{2}$}.

Combining with Eq.~(\ref{distance}) and substituting each
anisotropic model in the \textbf{$\widetilde{\chi}^{2}$}, one can
easily calculate the likelihood function of each parameter by
performing the Markov Chain Monte Carlo analysis. The
parameters to be constrained are $(g_0,\, \theta,\, \phi)$ and
$(g_0,\, g_1,\, \theta,\, \phi)$, respectively, where
($\theta,\,\phi$) is the direction of modulation. We then convert
($\theta,\,\phi$) into galactic coordinate ($l,\,b$).


\section{\normalsize{Redshift Tomography}}
\renewcommand{\thesection}{\arabic{section}}

\begin{figure*}
\begin{center}
 \includegraphics[width=5.5in]{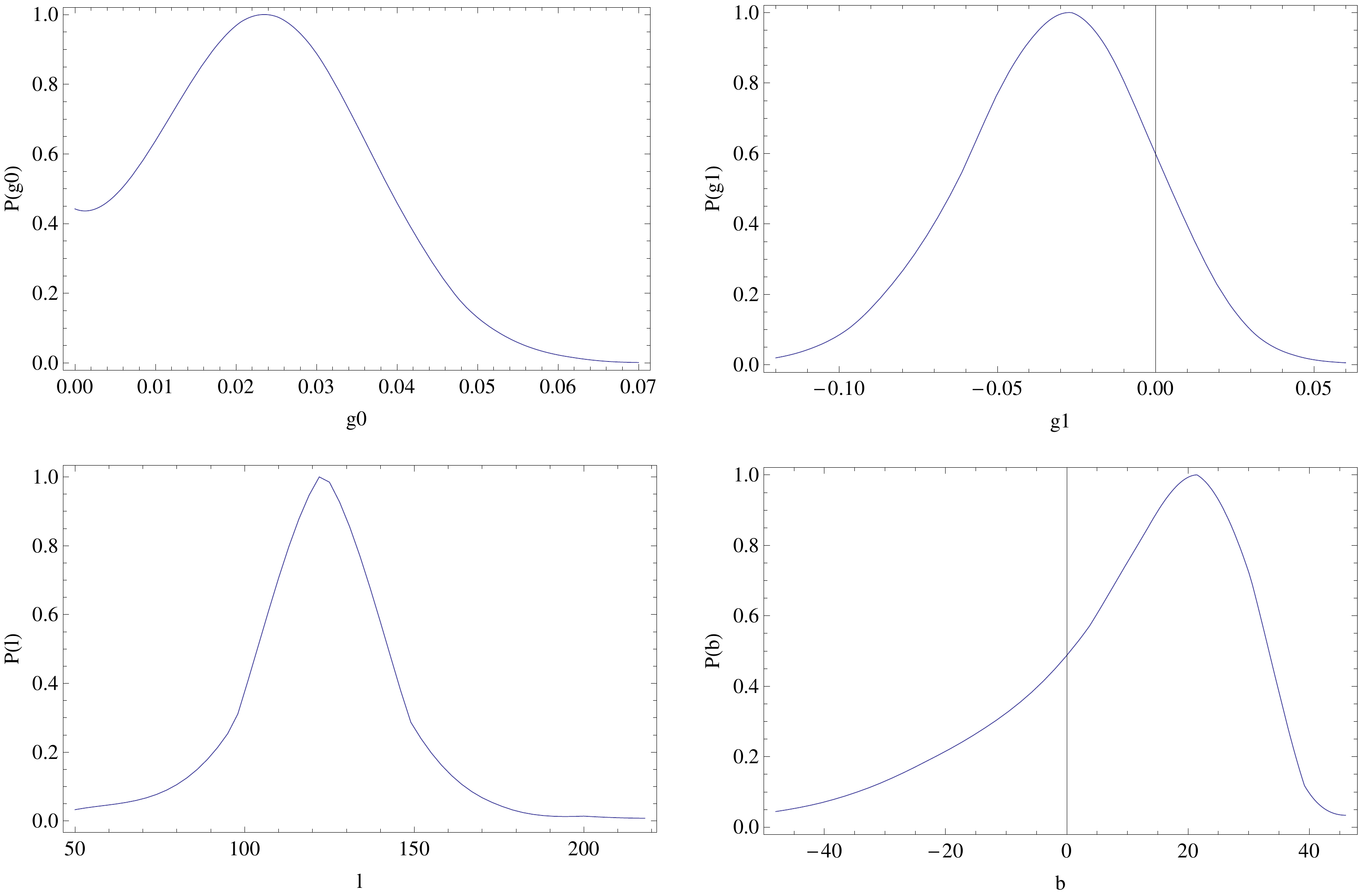}
 \caption{\label{fig:Likelihood}Likelihoods for parameters $(g_0, g_1, l, b)$ using
 the full Union2 data, with the
 best-fitting parameters $(g_0=0.030,\,g_1=-0.031,\,l=126^{\circ},\,b=13^{\circ})$.}
 \end{center}
\end{figure*}

\begin{center}
 \begin{figure}
 \includegraphics[width=3in]{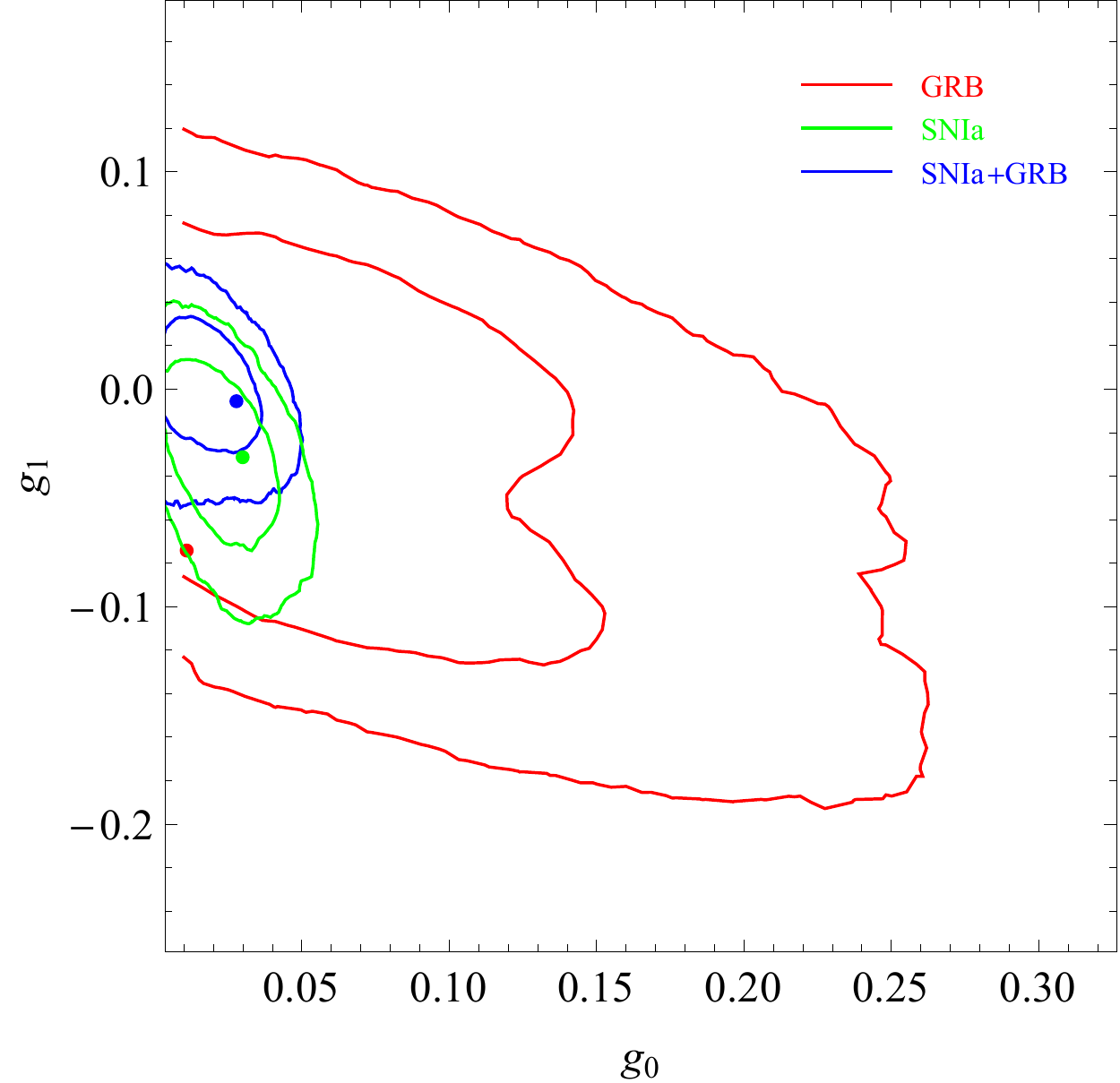}
 \caption{\label{fig:contour}68\% and 95\% joint
posterior probability distribution of parameters $(g_0, g_1)$. The red, green and blue contours represent the results
 using the GRB, SNIa and SNIa+GRB data, respectively.}

\end{figure}
 \end{center}

\begin{table*}[!h]
\begin{tabular}{|c|c|c|c|c|}
\hline ${\rm data\, sets}$  & $l[\rm degree]$  & $b[\rm
degree]$ & $g_0$  \tabularnewline \hline
\hline ${\rm
SNIa}$ & $128^{-23-94}_{+22+264}$ & $16^{-14-97}_{+21+64}$ &
$0.024^{-0.024-0.024}_{+0.008+0.021}$\tabularnewline \hline \hline${\rm
GRB}$ & $149^{-45-116}_{+35+243}$ & $-10^{-23-64}_{+46+81}$ &
$0.115^{-0.115-0.115}_{+0.036+0.082}$ \tabularnewline \hline
\hline${\rm SNIa+GRB}$ & $131^{-21-96}_{+13+256}$ &
$12^{-8-75}_{+18+47}$ & $0.027^{-0.027-0.027}_{+0.009+0.016}$\tabularnewline \hline
\end{tabular}
\tabcolsep 0pt \caption{\label{tab:g1}Constraints of the
directions and amplitude of maximum anisotropy for the constant modulation, using SNIa data, GRB data and
SNIa+GRB data, respectively. The error-bars quoted are $1\sigma$ and $2\sigma$
errors.} \vspace*{5pt}
\end{table*}

\begin{table*}[!h]
\begin{tabular}{|c|c|c|c|c|c|}
\hline ${\rm redshift\, range}$  & $l[\rm degree]$  & $b[\rm
degree]$ & $g_0$ & $g_1$ \tabularnewline \hline \hline $0-0.2$ &
$134^{-23-91}_{+13+212}$ & $4^{-7-59}_{+30+41}$ &
$0.045^{-0.044-0.045}_{+0.026+0.063}$ &
$-0.289^{-0.713-1.648}_{+0.410+1.013}$\tabularnewline \hline
\hline $0-0.4$ & $125^{-18-60}_{+17+255}$ & $8^{-7-88}_{+24+35}$ &
$0.036^{-0.028-0.036}_{+0.016+0.025}$ &
$-0.093^{-0.107-0.251}_{+0.163+0.354}$\tabularnewline \hline
\hline  $0-0.6$ & $121^{-18-81}_{+17+255}$ & $10^{-7-88}_{+24+35}$
& $0.033^{-0.033-0.033}_{+0.018+0.026}$ &
$-0.077^{-0.039-0.075}_{+0.099+0.163}$\tabularnewline \hline
\hline $0-0.8$ & $128^{-22-88}_{+17+265}$ &
$15^{-25-100}_{+19+63}$ & $0.030^{-0.030-0.030}_{+0.013+0.024}$ &
$-0.022^{-0.071-0.114}_{+0.064+0.185}$\tabularnewline \hline
\hline $0-1.0$ & $123^{-27-89}_{+20+269}$ & $14^{-22-79}_{+20+55}$
& $0.031^{-0.031-0.031}_{+0.012+0.024}$ &
$-0.040^{-0.050-0.097}_{+0.056+0.092}$\tabularnewline \hline
\hline $0-1.2$ & $125^{-33-90}_{+19+265}$ & $12^{-29-71}_{+20+43}$
& $0.032^{-0.032-0.032}_{+0.013+0.032}$ &
$-0.042^{-0.047-0.086}_{+0.050+0.082}$\tabularnewline \hline
\hline $0-1.4$ & $126^{-26-92}_{+17+266}$ & $13^{-25-82}_{+19+39}$
& $0.030^{-0.030-0.030}_{+0.010+0.021}$ &
$-0.031^{-0.039-0.075}_{+0.042+0.069}$\tabularnewline \hline
\end{tabular}
\tabcolsep 0pt \caption{\label{tab:tomography}Constraints of the
directions and amplitude of maximum anisotropy for different
redshift bins of the Union2 data. The error-bars quoted are
$1\sigma$ and $2\sigma$ errors.} \vspace*{5pt}
\end{table*}

\begin{table*}[!h]
\begin{tabular}{|c|c|c|c|c|c|}
\hline ${\rm data\, sets}$  & $l[\rm degree]$  & $b[\rm
degree]$ & $g_0$ & $g_1$ \tabularnewline
 \hline
\hline ${\rm SNIa}$ & $126^{-26-92}_{+17+266}$ & $13^{-25-82}_{+19+39}$
& $0.030^{-0.030-0.030}_{+0.010+0.021}$ &
$-0.031^{-0.039-0.075}_{+0.042+0.069}$ \tabularnewline
\hline \hline${\rm
GRB}$ & $336^{-223-303}_{+33+56}$ & $-5^{-26-50}_{+34+56}$ &
$0.011^{-0.011-0.011}_{+0.113+0.243}$ &
$-0.074^{-0.062-0.095}_{+0.137+0.170}$\tabularnewline \hline
\hline${\rm SNIa+GRB}$ & $129^{-23-96}_{+16+293}$ &
$16^{-10-89}_{+17+54}$ & $0.028^{-0.028-0.028}_{+0.013+0.021}$ &
$-0.006^{-0.034-0.055}_{+0.025+0.052}$\tabularnewline \hline
\end{tabular}
\tabcolsep 0pt
\caption{\label{tab:g3}Constraints of the
directions and amplitude of maximum anisotropy for the linear modulation, using SNIa, GRB data and
SNIa+GRB data, respectively. The error-bars quoted are $1\sigma$ and $2\sigma$
errors.} \vspace*{5pt}
\end{table*}

\begin{table*}[!h]
\begin{tabular}{|c|c|c|c|c|c|}
\hline ${\rm redshift\, range}$  & $l[\rm degree]$  & $b[\rm
degree]$ & $g_0$ & $g_1$ \tabularnewline \hline \hline $0-0.2$ &
$134^{-23-91}_{+13+212}$ & $4^{-7-59}_{+30+41}$ &
$0.045^{-0.044-0.045}_{+0.026+0.063}$ &
$-0.289^{-0.713-1.648}_{+0.410+1.013}$\tabularnewline \hline
\hline $0.2-0.4$ & $253^{-167-220}_{+107+138}$ &
$-49^{-4-34}_{+89+133}$ & $0.086^{-0.086-0.086}_{+0.287+0.552}$ &
$-0.309^{-1.073-1.670}_{+0.439+0.616}$\tabularnewline \hline
\hline  $0.4-0.6$ & $334^{-279-301}_{+56+59}$ &
$-18^{-24-58}_{+53+108}$ & $0.487^{-0.487-0.487}_{+0.412+0.511}$ &
$-0.077^{-0.039-0.075}_{+0.099+0.163}$\tabularnewline \hline
\hline $0.6-0.8$ & $149^{-115-116}_{+241+244}$ &
$63^{-106-140}_{+10+25}$ & $0.163^{-0.163-0.163}_{+0.787+0.833}$ &
$-0.191^{-1.138-1.257}_{+0.270+0.330}$\tabularnewline \hline
\hline $0.8-1.0$ & $146^{-96-113}_{+233+246}$ &
$-6^{-33-68}_{+46+90}$ & $0.997^{-0.996-0.997}_{+0.003+0.003}$ &
$-1.111^{-0.293-0.293}_{+1.355+1.355}$\tabularnewline \hline
\end{tabular}
\tabcolsep 0pt \caption{\label{tab:tomography2}Constraints on
direction and strength of modulation for several redshift bins of
the SNIa data fitting with the $\Lambda$CDM model, together with
the $1\sigma$ and $2\sigma$ errors of parameters $(l,\, b,\,
g_0,\, g_1)$.} \vspace*{5pt}
\end{table*}

\begin{center}
 \begin{figure}
 \includegraphics[width=3in]{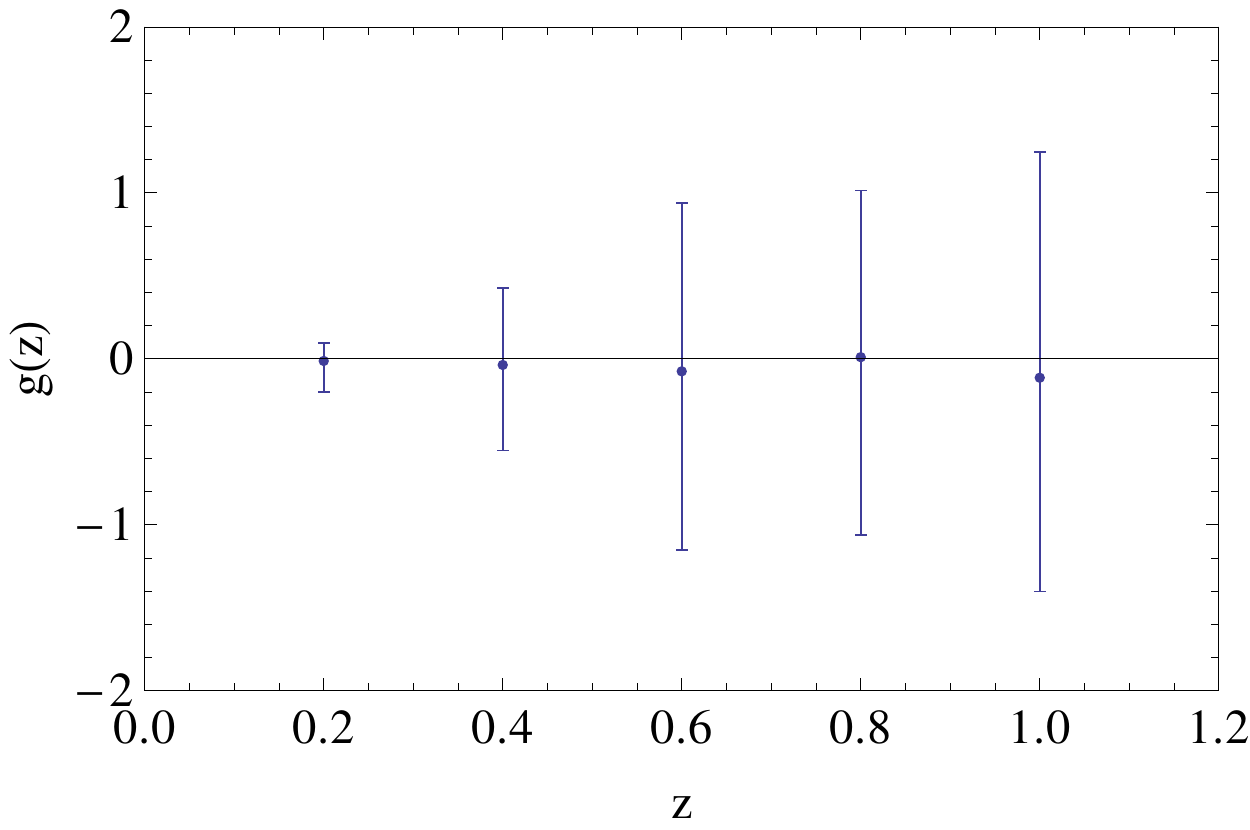}
 \caption{\label{fig:errorbar1} Best-fitting $g(z)$ with $1\sigma$ error at different
 redshifts.}

\end{figure}
 \end{center}

 \begin{table*}[!h]
\begin{tabular}{|c|c|c|c|c|c|}
\hline ${\rm data\, sets}$ & $l[\rm degree]$  & $b[\rm degree]$ &
$g_0$ & $g_1$ \tabularnewline \hline \hline
${\rm SNIa}$ &$128^{-28-88}_{+17+256}$ & $14^{-30-74}_{+17+39}$ &
$0.034^{-0.033-0.034}_{+0.013+0.026}$ &
$-0.045^{-0.056-0.093}_{+0.049+0.090}$\tabularnewline \hline
${\rm GRB}$ &$340^{-227-306}_{+35+53}$ & $-4^{-30-54}_{+34+60}$ &
$0.005^{-0.005-0.005}_{+0.127+0.217}$ &
$-0.027^{-0.109-0.112}_{+0.072+0.072}$\tabularnewline \hline
${\rm SNIa+GRB}$ &$129^{-21-95}_{+17+264}$ & $15^{-11-73}_{+18+35}$ &
$0.028^{-0.028-0.028}_{+0.012+0.022}$ &
$-0.006^{-0.033-0.048}_{+0.026+0.049}$\tabularnewline \hline
\end{tabular}
\tabcolsep 0pt \caption{\label{tab:model1}Directions of maximum
anisotropy fitting with the $w$CDM model, together
with the $1\sigma$ and $2\sigma$ errors of parameters $(l,\, b,\,
g_0,\, g_1)$, using the SNIa data and GRB data.} \vspace*{5pt}
\end{table*}

\begin{table*}[!h]
\begin{tabular}{|c|c|c|c|c|c|}
\hline ${\rm data\, sets}$ & $l[\rm degree]$  & $b[\rm degree]$ &
$g_0$ & $g_1$ \tabularnewline \hline \hline
${\rm SNIa}$ &$127^{-21-80}_{+17+230}$ & $15^{-32-77}_{+17+61}$ &
$0.035^{-0.035-0.035}_{+0.015+0.027}$ &
$-0.049^{-0.046-0.347}_{+0.049+0.102}$\tabularnewline \hline
${\rm GRB}$ &$339^{-224-306}_{+26+54}$ & $-6^{-25-46}_{+34+54}$ &
$0.008^{-0.008-0.008}_{+0.131+0.239}$ &
$-0.073^{-0.069-0.104}_{+0.141+0.218}$\tabularnewline \hline
${\rm SNIa+GRB}$ &$131^{-22-96}_{+14+262}$ & $16^{-10-77}_{+15+62}$ &
$0.028^{-0.028-0.028}_{+0.010+0.023}$ &
$-0.006^{-0.032-0.050}_{+0.025+0.045}$\tabularnewline \hline
\end{tabular}
\tabcolsep 0pt \caption{\label{tab:model2}Directions of maximum
anisotropy fitting with the CPL model, together
with the $1\sigma$ and $2\sigma$ errors of parameters $(l,\, b,\,
g_0,\, g_1)$, using the SNIa data and GRB data.} \vspace*{5pt}
\end{table*}

\begin{table}[!h]
\begin{tabular}{|c|c|c|c|}
\hline ${\rm Models}$  & ${\rm Parameters}$  & ${\rm
Bayes~factor}$ & ${\rm Comparison}$ \tabularnewline \hline \hline
$\Lambda$CDM & $\Omega_{\textrm{m0}}$ & $-0.35$ & ${\rm Weak}$
\tabularnewline \hline \hline $w$CDM & $\Omega_{\textrm{m0}},\,w$
& $1.08$ & ${\rm Significant}$\tabularnewline \hline \hline CPL &
$\Omega_{\textrm{m0}},\,w_0,\,w_1$ & $3.17$ & ${\rm
Strong~to~very~strong}$\tabularnewline \hline
\end{tabular}
\tabcolsep 0pt \caption{\label{tab:bayes}Results of comparison by
using the Bayes factor for anisotropic and isotropic dark anergy
models.} \vspace*{5pt}
\end{table}

Following the procedure introduced in Sec.2, one can obtain the
best-fitting values and errors of parameters by performing the
Markov Chain Monte Carlo analysis in the multidimensional
parameter space. During the procedure, we use the best-fitting
values obtained by \textit{WMAP} group to calculate $d_{\rm
L}^0(z)$ in the isotropic background. We choose
$\Omega_{\textrm{m0}}=0.274$ for $\Lambda$CDM model,
$\Omega_{\textrm{m0}}=0.274,\,w=-1.037$ for $w$CDM model, and
$\Omega_{\textrm{m0}}=0.274,\,w_0=-1.17,\,w_1=0.35$ for CPL
parametrization, which are obtained by using
\textit{WMAP}+eCMB+BAO+$H_0$+SNe data sets~\cite{Hinshaw:2012fq}.

First, we consider the constant modulation function $g(z)=g_0$. By
using the full Union2 data, we obtain the results listed in
Table~\ref{tab:g1}, which shows that the maximum deviation from
the $\Lambda$CDM model is 0.024, and the maximum anisotropy
direction is ($\sim(l,\,b)=(128^{\circ},\,16^{\circ}$), but we can
not exclude the case $g_0=0$ at $1\sigma$ confidence level.
We also use the GRB data to explore the high-redshift
behavior, which can extend our detection to $z=6.6$. In order to
check the consistency between the SNIa data and GRB data, we show
the fitting results by using GRB data, SNIa data and SNIa+GRB data
in Table~\ref{tab:g1}. The result shows that GRB data prefer
larger longitude and negative latitude, compared with the results
of SNIa data. But the maximum deviation axes are both consistent
within $1\sigma$ error, and the anisotropy level are both close to
0, which means that the GRB data and the Union2 data are
consistent on this issue.


Next, in order to explore the possible redshift dependence of the
anisotropy, we consider the linear modulation function and
implement a redshift tomography analysis, taking the same
procedure as before for the following redshift slices: 0-0.2,
0-0.4, 0-0.6, 0-0.8, 0-1.0, 0-1.2, 0-1.4. Our results for
$\Lambda$CDM model are summarized in Table \ref{tab:tomography}.

The redshift tomography analysis here shows that the preferred
axes at different redshifts are all located in a relatively small
region of the Galactic Hemisphere
($\sim(l,\,b)=(126^{\circ},\,13^{\circ})$ for Union2 data). Note
that the meaning of this best-fitting direction is the same as
$(l,\,b)=(306^{\circ},\,-13^{\circ})$, because both directions are
located on the same axis, and both are directions of maximum
deviation from the $\Lambda$CDM model. The best-fitting direction
is consistent with the result in Ref.~\cite{Mariano:2012wx} at
$1\sigma$ confidence level. We also re-examined the dark energy
dipole by using the Union2 data with the same method proposed in
Ref.~\cite{Mariano:2012wx}, showing that the dark energy dipole is
indeed aligned with the corresponding fine structure constant
cosmic dipole, where the dark energy dipole direction is
$(l=309.4^{\circ}\pm18.0^{\circ},\,b=-15.1^{\circ}\pm11.5^{\circ})$
and the fine structure dipole direction is
$(l=320.5^{\circ}\pm11.8^{\circ},\,b=-11.7^{\circ}\pm7.5^{\circ})$.

The effect of the GRB data and the consistency between the SNIa data and GRB data are shown in Table~\ref{tab:g3} for the linear
 modulation case. One can see from the table that the GRB data and the Union2 data are consistent at $1\sigma$ confidence level.

Using the full Union2 data, we plot the likelihood of the
parameters $(g_0,\,g_1,\,l,\,b)$ in Fig.~\ref{fig:Likelihood}. It
is obvious from Eq~(\ref{deviation}) that there are two maximum
anisotropic directions, which are settled on the same axis,
accordingly $g(z)$ can be positive or negative. We also
plot the joint posterior probability distribution of parameters
$(g_0, g_1)$ in Fig.~\ref{fig:contour}, showing that $g_0$ and
$g_1$ are negatively correlated and that the GRB data give a much weaker constraint compared with the SNIa data. Note that since we consider
dipole modulation ($s=1$ in Eq.~(\ref{deviation})), the positive
$g(z)$ at one direction is equivalent to the negative $g(z)$ at
opposite direction ($\hat{z}\rightarrow -\hat{z}$). Therefore we
restrict $g_0>0$ in our likelihood analysis.


Furthermore, we constrain the redshift dependence of the
anisotropy by using SNIa data located only in different redshift
bins: $0-0.2, 0.2-0.4, 0.4-0.6, 0.6-0.8, 0.8-1.0$, thus to avoid
influences from other bins. Accordingly, we obtain $N=220, 124,
102, 50$ and $41$ data points in different redshift bins,
respectively. Our results for $\Lambda$CDM model are summarized in
Table~\ref{tab:tomography2}. Fig.~~\ref{fig:errorbar1} shows the
best-fitting $g(z)$ with $1\sigma$ error at different redshifts.
It is clear that the anisotropy level $g(z)$ is close to 0, which
means that the $\Lambda$CDM model is still consistent with the
Union2 data very well. The error increases as redshift increases,
and we can not exclude the case $g(z)=0$ even at $68.3\%$
confidence level. Note that the anisotropic direction changes as
redshift bin changes, but theses directions are consistent with
each other at $68.3\%$ confidence level.

We also consider the $w$CDM and the CPL parameterized dark energy
models as the isotropic background. We use $w$ and ($w_0,\,w_1$)
as the isotropic background dark energy parameters and fit our
anisotropic parameters using Union2 and GRB data, and the
results are summarized in Table~\ref{tab:model1} and Table~\ref{tab:model2},
respectively,. The results of
constraints on $(l,\,b,\,g_0,\,g_1)$ are not much different from
the case of the $\Lambda$CDM model. This means that the
best-fitting value of the maximum deviation direction from the
isotropic background is not sensitive to the details of isotropic
dark energy models. In addition, we can also see that using the GRB data only one would obtain a larger longitude
and negative latitude, compared with the result using the SNIa data, but both data are still consistent at $1\sigma$
confidence level.

Finally, we make Bayesian statistical comparison between the
isotropic dark energy models and the anisotropic dark energy
models, in order to see which model is more favored by
observational data. The Bayesian Evidence $E$ provides us with a
good metric to quantify the level of consistency between each
model and observational data~\cite{Jeffreys},
\begin{equation}
E=\int
\textrm{Pr}(\mathbf{D}|\bf{\theta})\textrm{Pr}(\mathbf{\theta})d\mathbf{\theta},
\label{eq:bayes}
\end{equation}
where
$\textrm{Pr}(\mathbf{D}|\mathbf{\theta})=\mathcal{L}(\mathbf{\theta})$
is the likelihood function, $\textrm{Pr}(\mathbf{\theta})$ is the
prior distribution of the model parameter vector
$\mathbf{\theta}$, which is usually assumed to be uniform or
Gaussian. The logarithmic ratio of Evidences between the two
models $\log(B_{AB}) \equiv \log(E(A)/E(B))$, also known as the
Bayes factor, measures the goodness of fit of the models. In the
following, we calculate the Bayesian Evidence ratio of the
anisotropic model ($A$) and the isotropic model ($B$) by assuming
the multivariate Gaussian distribution of likelihood function and
the uniform distribution of the prior. According to the Jeffreys
grades \cite{Jeffreys}, the result is summarized as follows (see
Table~\ref{tab:bayes} as well).

\begin{itemize}
  \item [(1)] If the background is described by the $\Lambda$CDM model, the parameters for isotropic and anisotropic models are $(\Omega_{\textrm{m0}})$ and
  $(l,\, b,\, g_0,\, g_1)$, respectively. The Bayes factor is $-0.35$, which means that there is ``weak'' evidence that the $\Lambda$CDM model is better than the anisotropic model.
  \item [(2)] If the background is described by the $w$CDM model, the parameters for isotropic and anisotropic models are $(\Omega_{\textrm{m0}},\,w)$ and
  $(l,\, b,\, g_0,\, g_1)$, respectively. The Bayes factor is $1.08$, which means that there is a ``significant'' evidence that the anisotropic model is better than the $w$CDM model.
  \item [(3)] If the background is described by the CPL parametrization, the parameters for
isotropic and anisotropic models are
$(\Omega_{\textrm{m0}},\,w_0,\,w_1)$ and $(l,\, b,\, g_0,\, g_1)$,
respectively. The Bayes factor is $3.17$, which means that there
is ``strong to very strong'' evidence that the anisotropic dark
energy model is better than the CPL parametrization.
\end{itemize}

\subsection{\normalsize{Distinguish Anisotropic Dark Energy With Peculiar Velocity Field Model}}
In this paper, we consider the model of anisotropic dark energy, which may potentially be degenerated with
the peculiar velocity field model in an isotropic background. For example, the best-fitting bulk flow direction found in Ref.~\cite{Rathaus:2013ut}
is quite close to the direction $(l,\,b)=(306^{\circ},\,-13^{\circ})$.
This is because, the peculiar motion of galaxies or supernovae can
also produce a direction-dependent luminosity distance, which may indicate some preferred direction. With the current data at low redshift, we are not able to distinguish them, but we propose the following method which is useful for doing this in advance of new data.

First, if the anisotropy is caused by the peculiar velocity, the
anisotropic direction should be randomly distributed on different
cosmic scales, because peculiar velocity is driven by emergent of
large scale structure, but if the anisotropy is caused by the dark
energy dipole, the anisotropic direction should be a constant on
all cosmic scale, due to the non-local effect of dark energy. So
redshift tomography method may tell the differences between the
two models if high-$z$ SNIa data are available.

Second, peculiar velocity is a local effect and should be zero if averaged on the whole cosmic scale~\cite{Li:2012cv}. But the
dark energy dipole should not change with the redshift. So by average the galaxy luminosity over a large volume, one can distinguish
where the direction-dependence of luminosity is due to peculiar velocity or dark energy dipole.

In addition, the Integrated Sachs-Wolfe effect (ISW) can be used to detect the dark energy dipole, because
if the accelerating expansion is anisotropic, the photons that travel from different distances in different directions are potentially observable. On the other hand, peculiar motion of galaxies can only re-scatter the CMB photons and produces secondary anisotropic effect, which acquires its maximum at galaxy and cluster scales.

\subsection{\normalsize{Comparison With Other Methods}}
As discussed in Sec.1, many studies have been made on the
issue of cosmic anisotropy, and those results are compatible at
certain confidence level~\cite{Kalus:2012zu}, the anisotropic
directions are in the vicinity to the \textit{WMAP} cold spot
~\cite{Bennett:2010jb}.  Different from other methods, here we
used the luminosity distance as the diagnostics to search for the
maximum anisotropic direction, because anisotropy can directly
affect the expansion rate and lead to anisotropic luminosity
distance, no matter what it is caused physically.  In addition,
since the deviation from isotropy is tiny, we  therefore
parameterize the deviation as a linear function of redsfhift. This
method is simple but can be applied to analyze many different
anisotropic models.

For example, if the anisotropy caused by an anisotropic
equation of state of dark energy, the Hubble parameter shall also
be anisotropic, leading the luminosity distance to deviate from
the isotropic universe, which can be detected with our method. And
if the anisotropy is due to the anisotropic background geometry,
such as the Bianchi I model, the anisotropic scale factor will
lead the luminosity distance to be direction dependent.  By taking
the method we proposed here and using SNIa data in different sky
patches, one can also detect the scale factors in different
directions and quantify the anisotropic level.


\section{\normalsize{Conclusions}}
\renewcommand{\thesection}{\arabic{section}}

There is a tentative evidence that the anisotropic direction on the cosmic expansion exists. If such a cosmological
preferred axis indeed exists, one has to consider an
anisotropic expanding universe, instead of the
isotropic cosmological model.

In this paper, we investigated the plausible anisotropy in the
accelerating expansion universe with the Union2 data and high-$z$
GRB data. We constructed an anisotropic dark energy model, where
we quantified the strength of modulation as a constant over all
redshift and the direction of maximum expansion as $(l,\,b)$. By
using the full Union2 data, we found that the maximum deviation
from the $\Lambda$CDM universe is $0.024^{-0.024}_{+0.008}$, and
the maximum anisotropy direction is
$(l,b)=(128^{\circ},\,16^{\circ})$, but we can not exclude the
case $g_0=0$ at $1\sigma$ confidence level. We also
compared the results by using the GRB data and SNIa data, showing
that both data are consistent at $1\sigma$ confidence level, and
that the GRB data give a much weaker constraint compared with the
SNIa data.

If the modulation is a linear function $g(z)=g_0+g_1z$, we
found that the maximum anisotropic deviation direction is
$(l,b)=(126^{\circ},\,13^{\circ})$ (or equivalently
$(l,b)=(306^{\circ},\,-13^{\circ})$), and the maximum anisotropy
level is described by the parameters
$g_0=0.030_{+0.010}^{-0.030},\,g_1=-0.031_{+0.042}^{-0.039}$
(obtained using Union2 data, at $1\sigma$ confidence level).
Furthermore, we used the redshift tomography method by adding in
the GRB data, and we found that the anisotropy strength $g(z)$ is
close to 0 within $1\sigma$ confidence level, which indicates that
there is no strong evidence against isotropic $\Lambda$CDM model.

We also discussed the cases where the dark energy equation of
state is described by a constant $w$ and $w(z)=w_0+w_{1}z/(1+z)$,
respectively, but the results show a similar anisotropic
direction. This indicates that the best-fitting value of the
maximum deviation direction from the isotropic background is not
sensitive to the isotropic dark energy models.

Finally, by using the Bayesian Evidence, we found that the
anisotropic dark energy model does not show great statistical
evidence better than isotropic $w$CDM model, except that there is
a slightly greater evidence of anisotropic dark energy than CPL
isotropic dark energy model.


\begin{acknowledgments}
This work was supported in part by the National Natural Science Foundation of China
(No.10821504,No.10975168 and No.11035008), and in part by the Ministry of Science and
Technology of China under Grant No. 2010CB833004.

This publication was made possible through the support of a grant from the John Templeton Foundation and National Astronomical Observatories of Chinese Academy of Sciences. The opinions expressed in this publication are those of the  authors do not necessarily reflect the views of the John Templeton Foundation  or National Astronomical Observatories of Chinese Academy of Sciences. The  funds from John Templeton Foundation were awarded in a grant to The University  of Chicago which also managed the program in conjunction with National
Astronomical Observatories, Chinese Academy of Sciences.
\end{acknowledgments}


%
\begin{table*}
\begin{centering}
\begin{tabular}{|c|c|c|c|c|c|c|c|}
\hline ${\rm ID}$  & ${\rm z}$  & $\mu\pm\sigma_{\mu}$  & ${\rm
Equatorial~Coordinate}$ &${\rm ID}$  & ${\rm z}$  &
$\mu\pm\sigma_{\mu}$  & ${\rm
Equatorial~Coordinate}$\tabularnewline \hline $970228$ & $0.70$  &
$42.72\pm0.68$  & $5^h1^m57^s,11^{\circ}46.4'$ &$030329$ & $0.17$
& $39.73\pm0.29$ & $10^h44^m50^s,21^{\circ}31'$\tabularnewline
\hline $970508$ & $0.84$  & $43.76\pm0.35$  &
$6^h53^m28^s,79^{\circ}17.4'$ & $030429$ & $2.66$ & $46.61\pm0.53$
& $12^h13^m18^s,-20^{\circ}51.2'$\tabularnewline \hline $971214$ &
$3.42$  & $47.54\pm0.59$  & $11^h56^m30^s,65^{\circ}12'$ &$030528$
& $0.78$ & $44.31\pm0.54$ &
$17^h4^m2^s,-22^{\circ}39'$\tabularnewline \hline $980613$ &
$1.10$  & $44.75\pm1.22$  & $10^h17^m46^s,71^{\circ}29.9'$
&$040924$ & $0.86$ & $43.61\pm0.55$ &
$2^h6^m19^s,16^{\circ}1'$\tabularnewline \hline $980703$ & $0.97$
& $43.84\pm0.32$  & $23^h59^m7^s,8^{\circ}35.6'$ &$041006$ &
$0.71$ & $43.92\pm0.42$ & $0^h54^m53^s,1^{\circ}12'$
\tabularnewline \hline $990123$ & $1.61$ & $44.66\pm0.37$ &
$15^h48^m14^s,51^{\circ}31'$ &$050126$ & $1.29$ & $45.74\pm0.52$ &
$18^h32^m27^s,42^{\circ}23'$ \tabularnewline \hline $990506$ &
$1.31$ & $43.76\pm0.53$ & $11^h54^m41^s,-26^{\circ}45'$ &$050318$
& $1.44$ & $45.95\pm0.44$ &
$3^h18^m43^s,-46^{\circ}24.2'$\tabularnewline \hline $990510$ &
$1.62$ & $45.36\pm0.31$ & $13^h38^m51^s,-80^{\circ}32'$ &$050319$
& $3.24$ & $47.73\pm0.93$ &
$10^h16^m38^s,43^{\circ}34'$\tabularnewline \hline $990705$ &
$0.84$ & $43.40\pm0.38$ & $5^h9^m32^s,-72^{\circ}9'$ &$050401$ &
$2.90$ & $45.94\pm0.55$ &
$16^h31^m31^s,2^{\circ}11'$\tabularnewline \hline $990712$ &
$0.43$ & $41.76\pm0.44$ & $22^h31^m50^s,-73^{\circ}24'$ &$050406$
& $2.44$ & $48.03\pm0.70$ &
$2^h17^m43^s,-50^{\circ}10'$\tabularnewline \hline $991208$ &
$0.71$ & $41.65\pm0.65$ & $16^h33^m55^s,46^{\circ}26'$ &$050408$ &
$1.24$ & $45.09\pm0.72$ &
$12^h1^m55^s,10^{\circ}52'$\tabularnewline \hline $991216$ &
$1.02$ & $43.12\pm0.35$ & $5^h19^m31^s,11^{\circ}11'$ &$050502$ &
$3.79$ & $47.24\pm0.64$ &
$13^h29^m46^s,42^{\circ}40'$\tabularnewline \hline $000131$ &
$4.50$ & $47.14\pm0.68$ & $6^h13^m33^s,-51^{\circ}55.6'$ &$050505$
& $4.27$ & $48.49\pm0.59$ &
$9^h27^m8^s,30^{\circ}15'$\tabularnewline \hline $000210$ & $0.85$
& $42.27\pm0.65$ & $1^h59^m15^s,-40^{\circ}40'$ &$050525$ & $0.61$
& $43.28\pm0.37$ & $18^h32^m35^s,26^{\circ}20'$\tabularnewline
\hline $000911$ & $1.06$ & $44.27\pm0.67$ &
$2^h18^m42^s,7^{\circ}48'$ &$050603$ & $2.82$ & $44.66\pm0.58$ &
$2^h39^m55^s,-25^{\circ}10.9'$\tabularnewline \hline $000926$ &
$2.07$ & $45.09\pm0.68$ & $17^h4^m15^s,51^{\circ}46.6'$ &$050802$
& $1.71$ & $45.52\pm0.98$ &
$14^h37^m9^s,27^{\circ}48'$\tabularnewline \hline $010222$ &
$1.48$ & $44.62\pm0.29$ & $14^h52^m17^s,43^{\circ}2'$ &$050820$ &
$2.61$ & $46.27\pm0.59$ &
$22^h29^m36^s,19^{\circ}34.7'$\tabularnewline \hline $010921$ &
$0.45$ & $42.53\pm0.54$ & $22^h55^m35^s,40^{\circ}56'$ &$050824$ &
$0.83$ & $44.07\pm1.19$ &
$0^h48^m57^s,22^{\circ}36'$\tabularnewline \hline $011211$ &
$2.14$ & $45.53\pm0.43$ & $11^h15^m15^s,-21^{\circ}56'$ &$050904$
& $6.29$ & $49.27\pm0.47$ &
$0^h54^m41^s,14^{\circ}8.3'$\tabularnewline \hline $020124$ &
$3.20$ & $46.73\pm0.37$ & $9^h33^m8^s,-11^{\circ}35.6'$ &$050908$
& $3.35$ & $47.00\pm0.76$ &
$1^h21^m51^s,-12^{\circ}58'$\tabularnewline \hline $020405$ &
$0.70$ & $43.47\pm0.46$ & $5^h19^m31^s,11^{\circ}11'$ &$050922$ &
$2.20$ & $45.57\pm0.52$ &
$21^h9^m34^s,-8^{\circ}46.3'$\tabularnewline \hline $020813$ &
$1.25$ & $43.95\pm0.33$ & $19^h46^m38^s,-19^{\circ}35'$ &$051022$
& $0.80$ & $43.77\pm0.28$ &
$23^h56^m0^s,19^{\circ}36'$\tabularnewline \hline $020903$ &
$0.25$ & $42.23\pm1.16$ & $22^h49^m1^s,-20^{\circ}56'$ &$051109$ &
$2.35$ & $45.84\pm0.80$ & $22^h1^m11^s,40^{\circ}50.2'$
\tabularnewline \hline $021004$ & $2.32$ & $46.60\pm0.48$ &
$0^h27^m5^s,-19^{\circ}6.1'$ &$051111$ & $1.55$ & $44.54\pm0.60$ &
$23^h12^m38^s,18^{\circ}20'$\tabularnewline \hline $021211$ &
$1.01$ & $43.49\pm0.55$ & $8^h8^m55^s,6^{\circ}44'$ &$060108$ &
$2.03$ & $48.85\pm1.07$ & $9^h47^m58^s,31^{\circ}55'$
\tabularnewline \hline $030115$ & $2.50$ & $46.25\pm0.57$ &
$11^h18^m30^s,15^{\circ}2'$ &$060206$ & $4.05$ & $46.37\pm0.60$ &
$13^h31^m47^s,35^{\circ}4.5'$\tabularnewline \hline $030226$ &
$1.98$ & $46.50\pm0.40$ & $11^h33^m36^s,25^{\circ}49.9'$ &$060116$
& $6.60$ & $48.33\pm0.92$ &
$5^h38^m48^s,-5^{\circ}27'$\tabularnewline \hline $030323$ &
$3.37$ & $47.65\pm0.96$ & $11^h33^m36^s,25^{\circ}49.9'$ &$060124$
& $2.30$ & $46.78\pm0.38$ &
$5^h8^m10^s,69^{\circ}42.5'$\tabularnewline \hline $030328$ &
$1.52$ & $44.68\pm0.37$ & $12^h10^m46^s,-9^{\circ}22.5'$ &$060115$
& $3.53$ & $47.78\pm0.79$ &
$3^h36^m11^s,17^{\circ}20.6'$\tabularnewline \hline $060210$ &
$3.91$ & $48.59\pm0.47$ & $3^h50^m55^s,27^{\circ}2'$ &$060526$ &
$3.21$ & $47.17\pm0.41$ &
$15^h31^m23^s,0^{\circ}14.7'$\tabularnewline \hline $060223$ &
$4.41$ & $47.64\pm0.54$ & $3^h40^m52^s,-17^{\circ}8'$ &$060604$ &
$2.68$ & $46.23\pm0.57$ &
$22^h28^m58^s,-10^{\circ}54'$\tabularnewline \hline $060418$ &
$1.49$ & $45.00\pm0.51$ & $15^h45^m41^s,-3^{\circ}38.6'$ &$060605$
& $3.80$ & $47.04\pm0.68$ &
$21^h28^m35^s,-6^{\circ}4'$\tabularnewline \hline $060502$ &
$1.51$ & $44.90\pm0.62$ & $16^h3^m26^s,66^{\circ}36'$ &$060607$ &
$3.08$ & $46.24\pm0.55$ &
$21^h58^m51^s,-22^{\circ}30.3'$\tabularnewline \hline $060510$ &
$4.90$ & $48.60\pm0.93$ & $6^h23^m29^s,-1^{\circ}10'$
\tabularnewline \hline
\end{tabular}
\par\end{centering}
\caption{\label{tab:GRB} Selected $67$ GRB samples with positions
described in the equatorial coordinates (right ascension and
declination).}
\end{table*}

%

\end{document}